\documentclass[iop,twocolappendix]{emulateapj}
\usepackage{amsmath,mathtools}
\begin{document}
\title{GRMHD simulations of visibility amplitude variability for Event Horizon Telescope images of Sgr A*}
\author{Lia Medeiros\altaffilmark{1}$^,$\altaffilmark{2}, Chi-kwan Chan\altaffilmark{1}, Feryal \"Ozel\altaffilmark{1}, Dimitrios Psaltis\altaffilmark{1}, Junhan Kim\altaffilmark{1}, \\Daniel P. Marrone\altaffilmark{1}, and Aleksander S{\c a}dowski\altaffilmark{3}}

\altaffiltext{1}{Steward Observatory and Department of Astronomy, University of Arizona, 933 N. Cherry Ave., Tucson, AZ 85721}
\altaffiltext{2}{Department of Physics, Broida Hall, University of California Santa Barbara, Santa Barbara, CA 93106}
\altaffiltext{3}{MIT Kavli Institute for Astrophysics and Space Research, 77 Massachusetts Ave, Cambridge, MA 02139}

\begin{abstract}
The Event Horizon Telescope will generate horizon scale images of the black hole in 
the center of the Milky Way, Sgr A*. Image reconstruction using interferometric 
visibilities rests on the assumption of a stationary image. We explore the limitations 
of this assumption using high-cadence disk- and jet-dominated  GRMHD simulations of 
Sgr A*. We also employ analytic models that capture the basic characteristics of the 
images to understand the origin of the variability in the simulated visibility 
amplitudes. We find that, in all simulations, the visibility amplitudes for baselines oriented 
parallel and perpendicular to the spin axis of the black hole follow general trends that do not
depend strongly on accretion-flow properties. This suggests that fitting EHT observations
with simple geometric models may lead to a reasonably accurate determination of the orientation 
of the black-hole on the plane of the sky. However, in the disk-dominated models, the locations 
and depths of the minima in the visibility amplitudes are highly variable and are not related
simply to the size of the black-hole shadow. This suggests that using time-independent 
models to infer additional black-hole parameters, such as the shadow size or the spin magnitude, 
will be severely affected by the variability of the accretion flow.

\end{abstract}

\keywords{accretion, accretion disks --- black hole physics --- Galaxy: center --- radiative transfer}

\section{Introduction}
\begin{figure*}[t!]
\centering
\includegraphics[trim=24 24 24 24, width=7.5in]{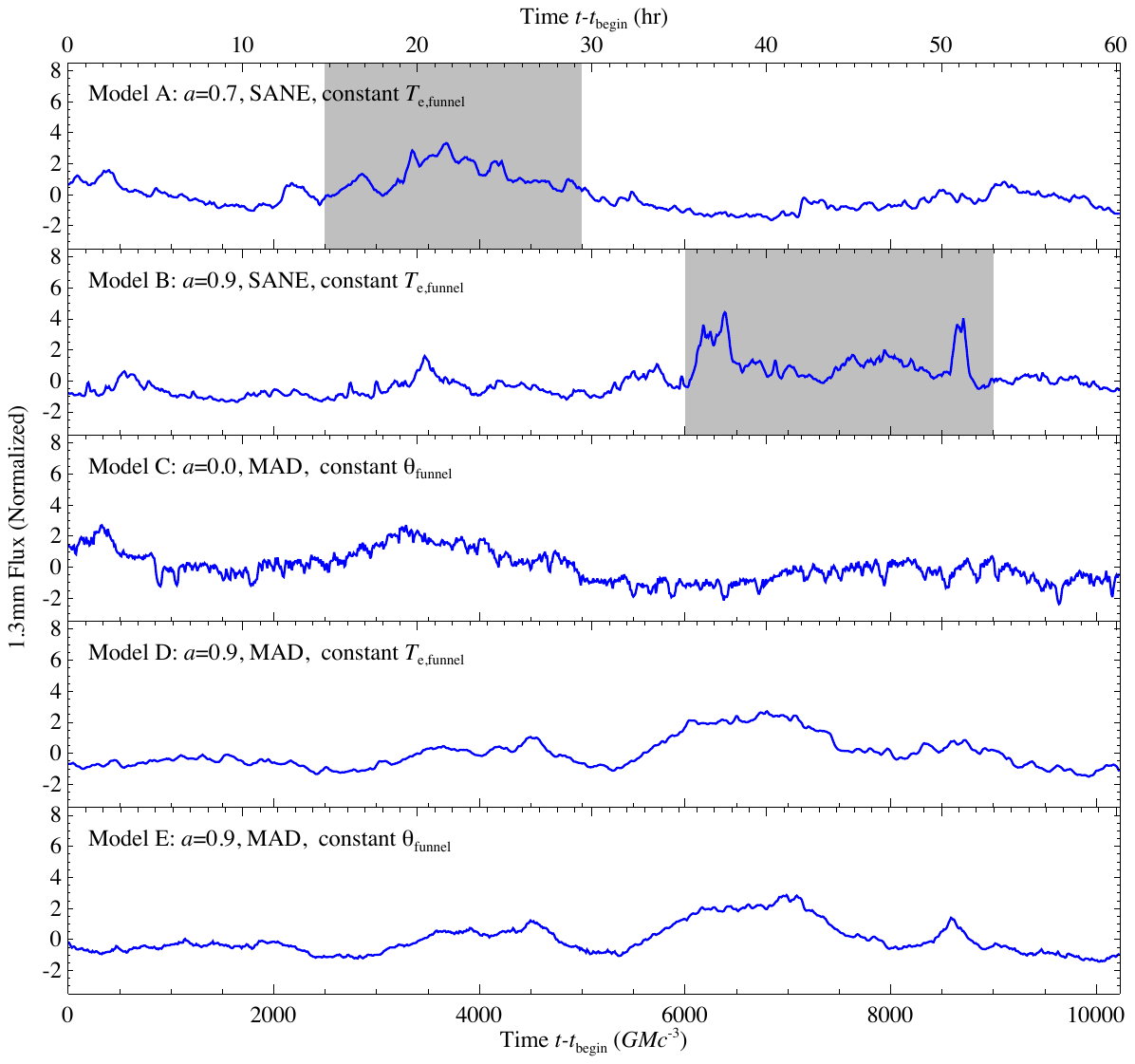}
\caption{The 1.3~mm flux, centered (subtracted by the mean) and normalized by the variance, as a function of time for the five simulations we explore in this paper (see \citealt{chan2015, 2015ApJ...799....1C}). Models A and B, the SANE models, have significant flaring events, shown in grey, which we have excluded from all calculations of averages throughout the paper. The models have a time resolution of $10\,M$, which is approximately equal to 212 s for the mass of Sgr~A$^*$ and a total duration of $\approx 60$ hours.}
\label{fig:lightcurve}
\end{figure*}

The Event Horizon Telescope (EHT) is a very long baseline interferometer (VLBI) with baselines ranging from Arizona to California and from the South Pole to Europe. It will measure interferometric visibilities and use them to generate, for the first time, black hole images with horizon scale resolution (see, e.g., \citealt{2009astro2010S..68D}). This will not only allow us to test theories of accretion physics but will also give us an unprecedented look at general relativistic effects in the strong field regime \citep{2011JPhCS.283a2030P}.  

Sagittarius A$^*$ (Sgr~A$^*$), the black hole at the center of our galaxy, is one of the primary targets for the EHT. It has a large angular size, a well constrained mass and distance \citep{2008ApJ...689.1044G, 2009ApJ...692.1075G}, and a broadband spectrum that has been studied for more than a decade (see \citealt{2001Natur.413...45B} and \citealt{2003Natur.425..934G} for early studies). The source-integrated monochromatic flux of Sgr~A$^*$ has been observed to be variable at many wavelengths on timescales of $\sim$hours including at 1.3~mm, the wavelength at which the EHT will operate \citep[e.g.,][]{2008ApJ...682..373M, 2008A&A...488..549P, 2009ApJ...691.1021D}. 
The EHT itself has also observed variability at 1.3~mm  \citep{2011ApJ...727L..36F}, indicating that Sgr~A$^*$ is structurally variable on scales of a few Schwarzschild radii.

Improvements in our theoretical understanding of black-hole accretion have led to a convergence in the properties of GRMHD simulations (see, e.g., \citealt{2009ApJ...706..497M,2010ApJ...716..504S,2010ApJ...717.1092D, 2012JPhCS.372a2023D,2012ApJ...755..133S,2013A&A...559L...3M,2009ApJ...701..521C, chan2015, 2015ApJ...799....1C}) and semi-analytic models inspired by them (e.g., \citealt{2006ApJ...636L.109B,2011ApJ...735..110B,2016ApJ...820..137B})
that can account for most observed characteristics of Sgr~A$^*$. A variety of these
models has been directly compared to early EHT data in order to perform parameter estimation and model comparison. 

In a previous set of papers, we reported a series GRMHD simulations with high 
spatial and time resolution \citep{chan2015, 2015ApJ...799....1C}.  We used these simulations first to study the variability as a function of wavelength for the source-integrated monochromatic flux of Sgr~A$^*$ and found two kinds of variability: long timescale flaring events and shorter timescale, persistent variability originating from the turbulent flow. Our disk dominated models are able to reproduce the flaring events observed from Sgr~A$^*$ at longer wavelengths but additional physics (such as non-thermal electrons) is required to reproduce X-ray flares (\citealt{2016ApJ...826...77B}; see also \citealt{2009ApJ...701..521C,2010ApJ...725..450D}).

Our simulations show that both long and short timescale variability occur at event horizon scales. 
The structure of the emission region, and therefore, the interferometric visibilities predicted by the simulations are highly variable. 
At event horizon scales, the dynamical timescale for Sgr A* is about 10 minutes, similar to the typical exposure time for the EHT. However, since the EHT is a VLBI instrument, it relies on the rotation of the Earth to trace baseline tracks in $u-v$ space. Therefore, during a single night of observation, different data points along baseline tracks correspond to different realizations of the turbulent structure of Sgr~A$^*$. 
Traditional image reconstruction algorithms rely on the assumption that the black hole image structure remains stationary on typical observation timescales, which range from about one hour for the shortest tracks to over ten hours for the longest. 

In order to account for the expected variability, a variable overall normalization of the image brightness between observing epochs was explicitly allowed for in some earlier studies but any variability on the image structure was assumed to be negligible (e.g., \citealt{2009ApJ...697...45B,2009ApJ...706..497M}). In other analyses, small scale Gaussian brightness fluctuations were added to smooth semi-analytical models (e.g.,~\citealt{2016ApJ...820..137B}) or the effects of image-structure variability on the parameter estimation were explored a posteriori (e.g., \citealt{2010ApJ...717.1092D}). Finally, \citet{2016ApJ...817..173L} explored methods for mitigating intrinsic source variability that introduces time variability to the image structure using one GRMHD simulation.

In this paper we take a step back and explore how the variability that we 
see in GRMHD simulations of accretion flows will be manifest on the $u-v$ plane. 
Specifically, we aim to characterize and understand the
variability in the location of salient features of black-hole images in
$u-v$ space, using these simulations. Our study allows us to explore the degree to which these salient features in the $u-v$ plane, which primarily determine the structures in the reconstructed images, evolve during the duration of an EHT imaging observation. It also helps us assess whether the location of such features can be used to measure fundamental properties of the black-hole that are fixed, such as the size of the black-hole shadow or the black-hole spin.

Because we want to quantify the time evolution of these features in the
$u-v$ plane as predicted by GRMHD simulations, we do not simulate 
particular EHT observations, consider the effects of the
Earth's rotation, or consider interstellar scattering. Of course, the
variability in the visibility amplitudes that the EHT will measure
will ultimately be a combination of the intrinsic source variability,
the effect of the Earth's rotation, and of the scattering screen. However, 
our goal is not to simulate a mock observation or perform a 
parameter estimation. We rather aim to test the degree to which one of the key assumptions 
in VLBI image reconstruction, i.e., that the images can be treated 
as stationary, will affect the interpretation of Sgr~A$^*$ observations.

To perform this analysis, we calculate and study the time-dependent 
visibilities at 1.3~mm for the five best-fit models of the GRMHD 
simulations that we reported earlier \citep{chan2015}. These models 
were chosen so that their average broadband spectra and 
image sizes at 1.3~mm match observations. 
We employ analytic models to understand the behavior and 
significance of various features in the visibility amplitudes and 
use our simulations to investigate the ability of analytic models 
to capture the gross features of the black-hole images. We explore 
the effects of intrinsic source variability on EHT closure phases 
in a separate paper \citep{2016arXiv161003505M}.

\section{The GRMHD+ray Tracing Simulations}

\begin{figure*}[t!]
\centering
\includegraphics[height=2in]{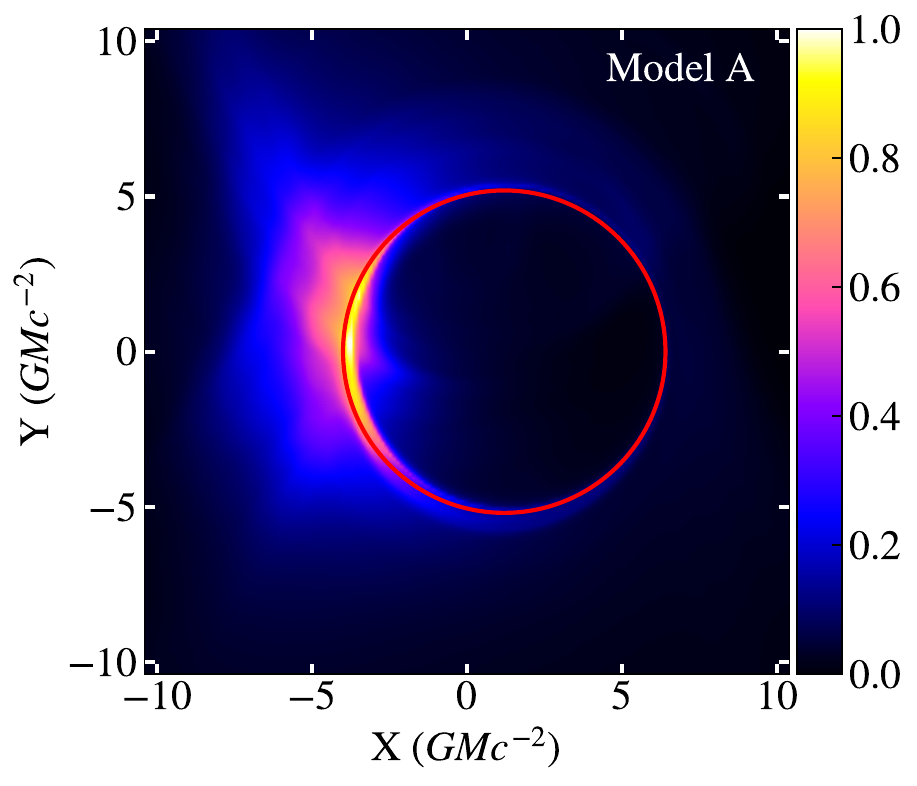}
\includegraphics[height=2in]{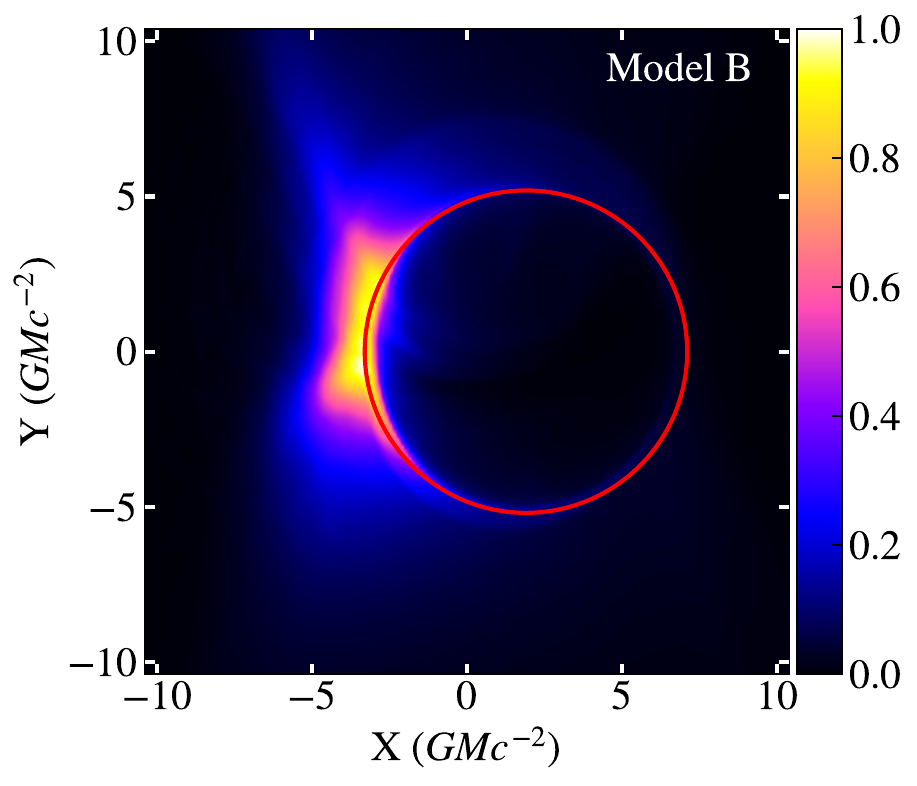}
\includegraphics[height=2in]{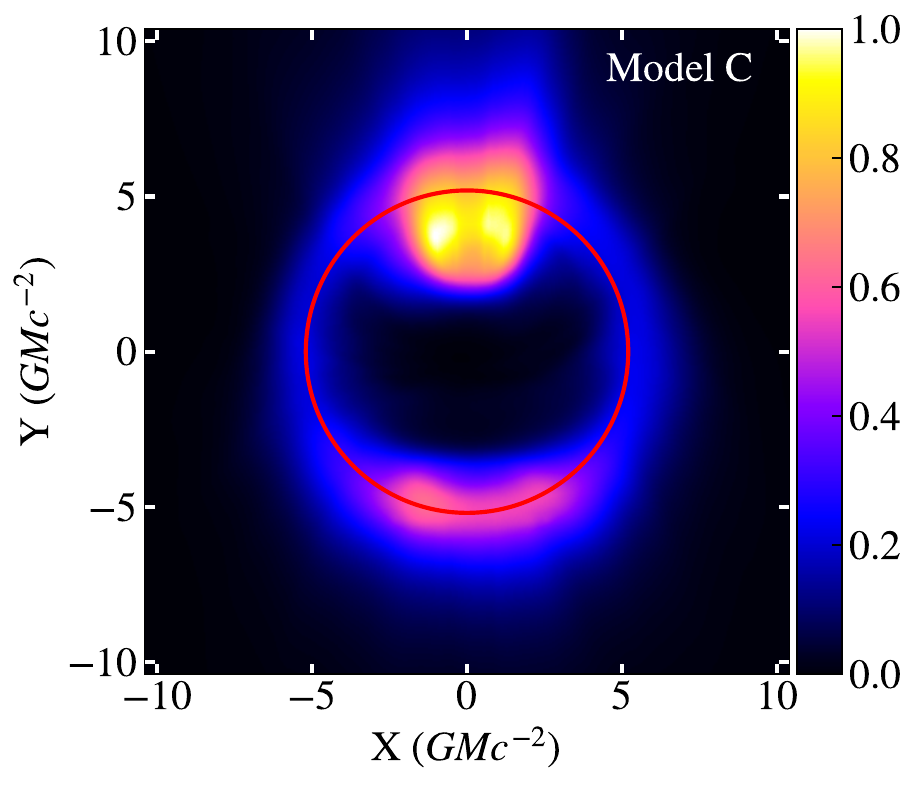}
\includegraphics[height=2in]{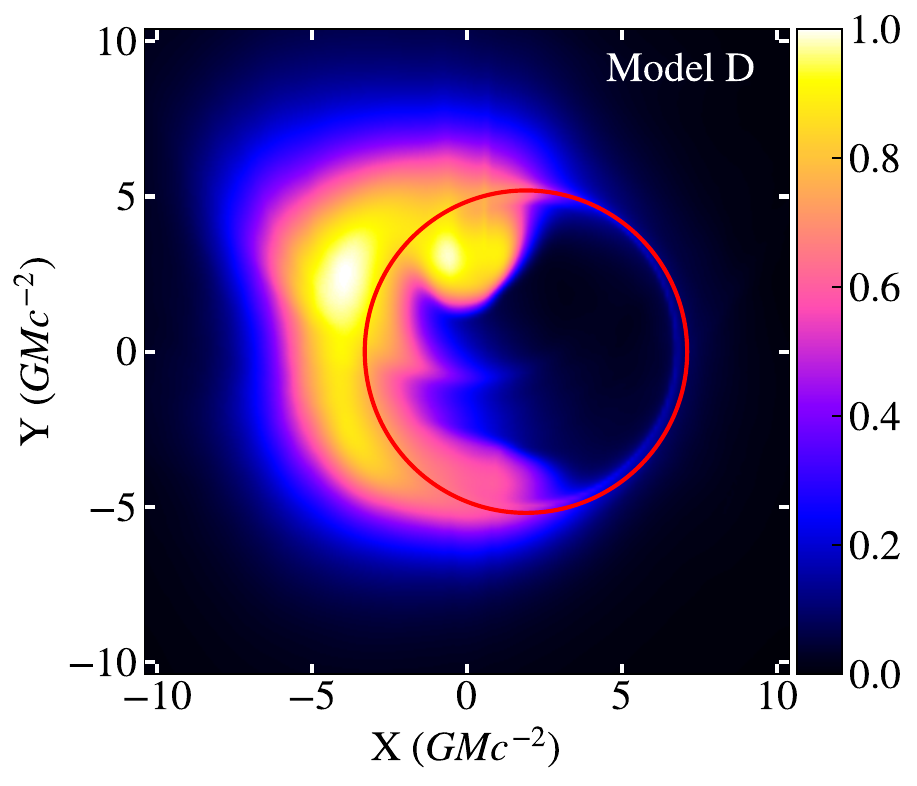}
\includegraphics[height=2in]{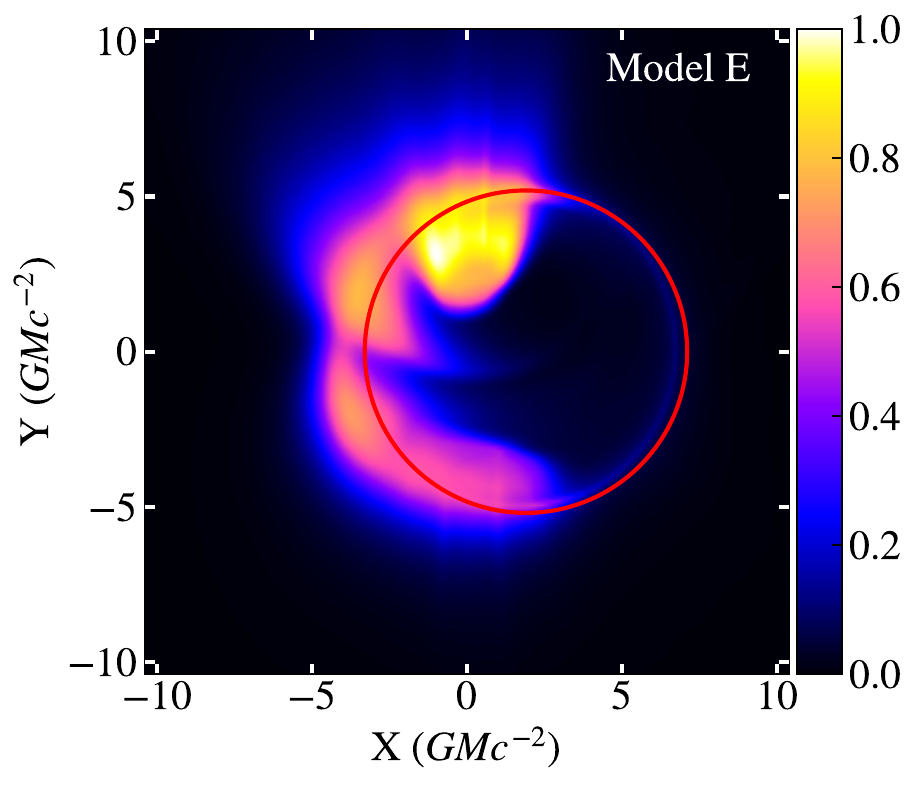}
\caption{The average 1.3~mm images of the five models we consider in this paper. The SANE models (A and B) have most of their emission originating from the disk region, while the MAD models (C, D, and E) have significant emission originating from the jets. Model C is unique, with a black hole spin of zero, which results in an almost symmetric image without the Doppler beaming effects that are present in the other models. The red circles indicate the expected size of the black hole shadow according to general relativity. The maximum intensity in each panel has been normalized to unity.}
\label{fig:average}
\end{figure*}

In previous papers, we explored a variety of GRMHD+ray tracing simulations for the accretion flow around Sgr~A$^*$ (see \citealt{2015ApJ...799....1C} and references therein for a detailed discussion of the simulation in the context of other related work). These 3 dimensional models were created using \texttt{HARM} \citep{ 2003ApJ...589..444G,sadowski_fix, 2013MNRAS.436.3856S} for the GRMHD simulations and \texttt{GRay} \citep{2013ApJ...777...13C} for solving the radiative transfer equation along null geodesics. We used simulations with a time resolution of $10\,GMc^{-3}$, equivalent to 212 s, where $G$ is the gravitational constant, $M = 4.3\times10^6\,M_\Sun$ is the mass of Sgr~A$^*$, and $c$ is the speed of light. For the remainder of this paper we employ gravitational units and set $G=c=1$. 

\texttt{GRay} employs the fast light approximation (see, e.g., \citealt{2009ApJS..184..387D,2011ApJ...735....9M,2012ApJ...744..187S} for models allowing for a finite speed of light) in which the speed of light is assumed infinite. At horizon scales, the light crossing time is $\sim 20$ s which is an order of magnitude smaller than our time resolution. 

We explored a large range of physical conditions by varying the black hole spin ($a$=0 to 0.9), the accretion rate, the magnetic field configuration, and the thermodynamic prescriptions for the electrons used in the plasma models. 
Specifically, we explored two different initial magnetic field configurations; the first consists of multiple small loops and  leads to weak, turbulent fields near the horizon and an emitting region at 1.3mm that is dominated by the disk region (SANE, Standard and Normal Evolution, see also, \citealt{2004ApJ...611..977M,2006ApJ...641..103H,2012ApJ...744..187S,2013A&A...559L...3M,2014A&A...570A...7M} for other SANE-like GRMHD simulations); the second consists of an initial magnetic field loop and leads to coherent magnetic field structures near the horizon and an emitting region at 1.3 mm that is dominated by the jets (MAD, Magnetically Arrested Disk, see, e.g., \citealt{2012JPhCS.372a2040T,2012MNRAS.423.3083M}).

Out of a large number of simulations, we identified five models which best fit the time averaged observations of Sgr~A$^*$  \citep{2015ApJ...799....1C}. 
The specific criteria we used to constrain these models are {\em (a)\/} a flux and a slope at $10^{11}$--$10^{12}$ Hz that matches observations, {\em (b)\/} a flux at $\simeq 10^{14}$ Hz that falls within the observed range of the highly variable infrared flux, {\em (c)\/} an X-ray flux that is consistent with $10\%$ of the observed quiescent flux, i.e., the percentage which has been attributed to emission from the inner accretion flow \citep{2013ApJ...774...42N}, and {\em (d)\/} a size of the emission region that is consistent with the size determined by the early EHT observations  \citep{2008Natur.455...78D}. 
All models have an observer inclination of i = 60$^{\circ}$ with respect to the spin axis of the black hole.

\begin{figure*}[t!]
\centering
\includegraphics[height=1.98in]{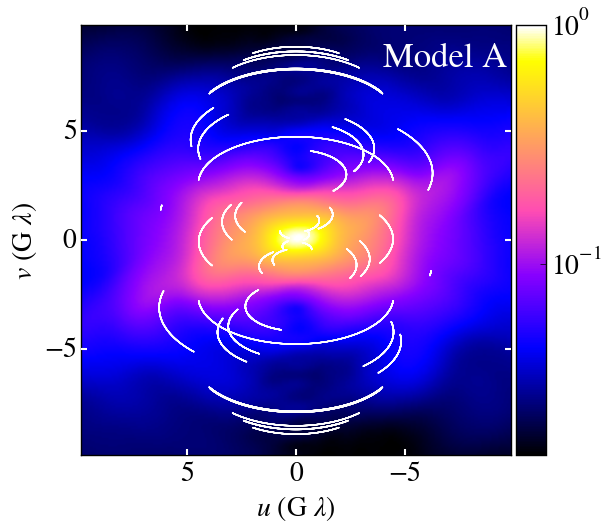}
\includegraphics[height=1.98in]{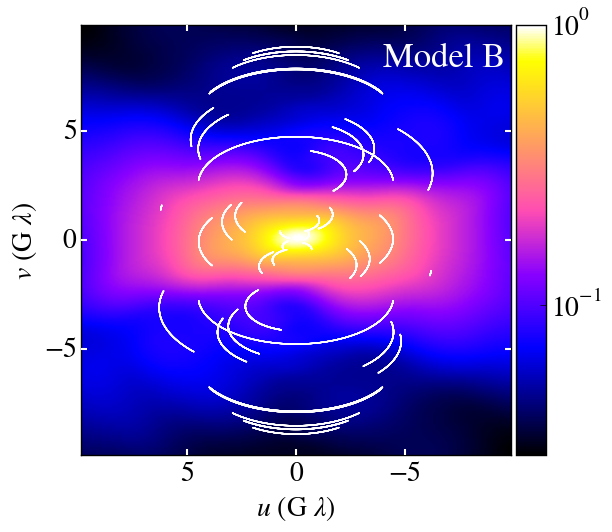}
\includegraphics[height=1.98in]{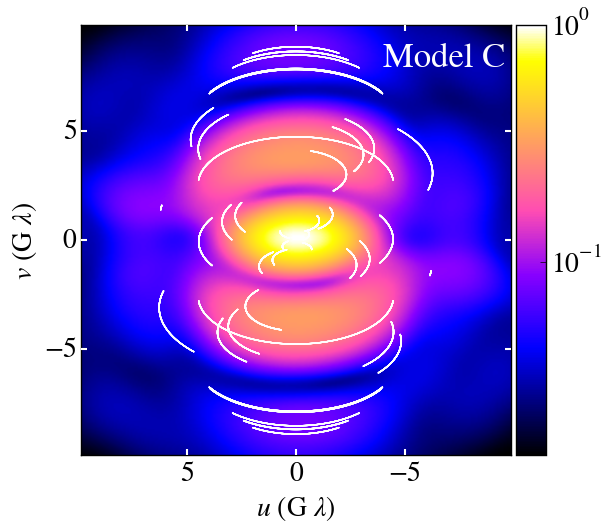}
\includegraphics[height=1.98in]{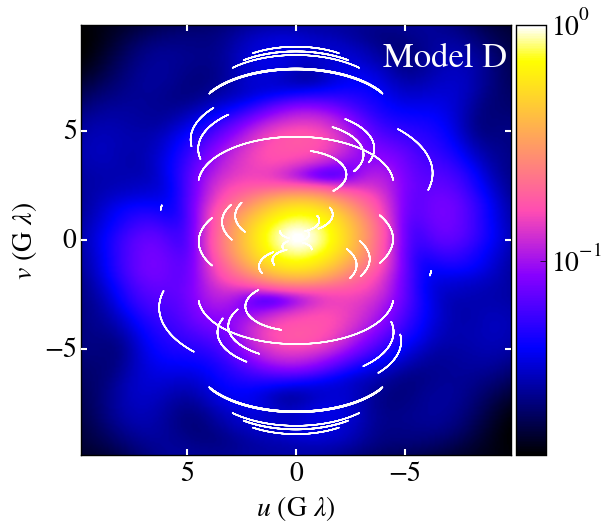}
\includegraphics[height=1.98in]{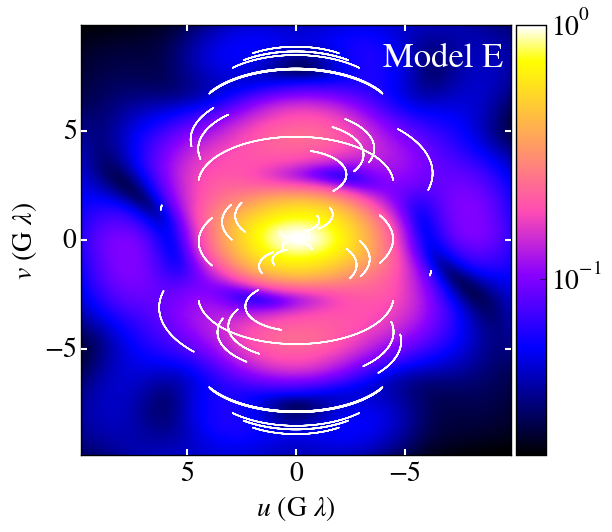}
\caption{The average 1.3~mm visibility amplitudes, calculated by taking the magnitudes of the complex Fourier components of the two dimensional Fourier transforms, of each snapshot of the five simulations shown in Figure~\ref{fig:lightcurve} and then averaging over the snapshots. 
The white lines denote the current and future tracks of the EHT baselines as the Earth rotates. Baselines are shown for an arbitrary N-S black hole orientation for illustrative purposes only.
The maximum visibility amplitude in each panel has been normalized to unity. 
\\
\\}
\label{fig:averageM}
\end{figure*}

Figure~\ref{fig:lightcurve} shows the light curves for the five models we consider in this paper. These light curves were calculated by computing the source integrated monochromatic flux from each model for 1024 late-time snapshots for a total duration that corresponds to 60 hrs for the assumed black hole mass. Models A and B have SANE flows and the same plasma model but differ on the choice of black hole spin: Model A has $a= 0.7$, while Model B has $a=0.9$. Both models A and B show large amplitude, short timescale variability that was shown to be consistent with broadband observations in \citet{chan2015}. Models D and E have the same black hole spin of $a=0.9$ and the same MAD configuration, but differ on the choice of the plasma model used.
Both models D and E show only long timescale, low amplitude variability. Model C consists of a MAD flow and a black hole spin of $a=0$. Model C shows very fast, quasi-periodic, low amplitude variability. 

In order to focus on the persistent variability from the turbulent flow, we ignore hereafter the intervals with large flaring events, shown in grey in Figure~\ref{fig:lightcurve}. 
These events were already discussed in \citet{chan2015}.   

Figure~\ref{fig:average} shows the average 1.3~mm images for the five models we consider. 
The SANE models (A and B) show a mostly connected emission region and appear to have their emission dominated by the disk. 
Models C and E on the other hand show disconnected emission regions and have most of their emission originating from the bases of the jets. 
Model D is a hybrid with a bright peak at the base of the jet (as in Models C and E) as well as a crescent shape (as in A and B). 
The red circles in the figures have a radius of $\sqrt{27}M$, i.e., the expected shadow radius of a non-rotating black hole according to general relativity. 
The images of the black hole shadow for models A, B, D, and E appear shifted to the right and have most of their emission originating from the left side due to Doppler beaming in a Kerr metric. 
The black hole shadow for model C is not shifted from the center and is much more symmetric than the others because the black hole spin in this model is zero. 

Although Figure~\ref{fig:average} shows resolved images of Sgr~A$^*$ at 1.3~mm, the EHT is an interferometer and will measure the complex Fourier components of Figure~\ref{fig:average}. 
To explore the properties of the actual observables, we show in Figure~\ref{fig:averageM} the average 1.3~mm visibility amplitudes for the five models.
We calculated these visibility amplitudes by performing, on each snapshot, the two dimensional Fourier transform
\begin{equation}
V(u,v) = \iint  I(\alpha,\beta) e^{-2 \pi i (u \alpha+v \beta)} d\alpha d\beta,
\end{equation}
where $\alpha \equiv X/D$, $\beta \equiv Y/D$, and $D$ is the distance to Sgr~A$^*$,
and then taking the mean magnitude of the complex Fourier components. 
The white lines are the current and planned tracks of the various baselines of the EHT shown for a particular N-S orientation of the black hole (see, e.g., \citealt{2009astro2010S..68D}). 
The visibility amplitude maps of models A and B appear elongated in the horizontal direction. 
This is because the Fourier transform is the conjugate of the original image and the original image is a crescent elongated in the vertical direction. 
Models C, D, and E show multiple emission peaks in the original image along the vertical direction.
This results in multiple peaks in the visibility amplitude maps, along the same axis.
Model C appears to be more symmetric in both the original image and its transform.

\section{Time Dependence of Visibilities}
Even though the visibility maps shown in Figure~\ref{fig:averageM} allow us to identify the gross features of the images, they do not faithfully represent the observations that the EHT will obtain.
As discussed in the introduction, the EHT relies on the rotation of the Earth to increase its coverage of the $u-v$ plane and, therefore, make a better image. 
However, Sgr~A$^*$ is variable on timescales that are much shorter than a day.
Since the EHT will observe Sgr~A$^*$ for multiple days each year for multiple years, it will measure a distribution of data points at each $u-v$ point along the baseline tracks. 
We now aim to quantify the effect of variability on the structure of the visibility amplitude.

\begin{figure*}[t!]
\centering
\includegraphics[height=2in]{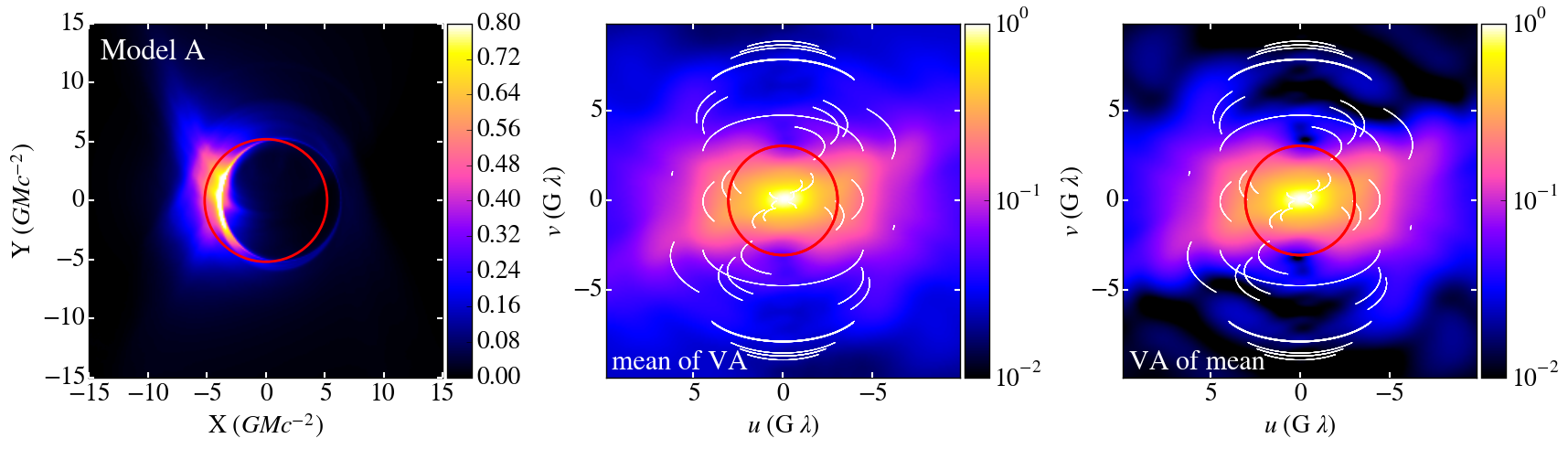}
\includegraphics[height=2in]{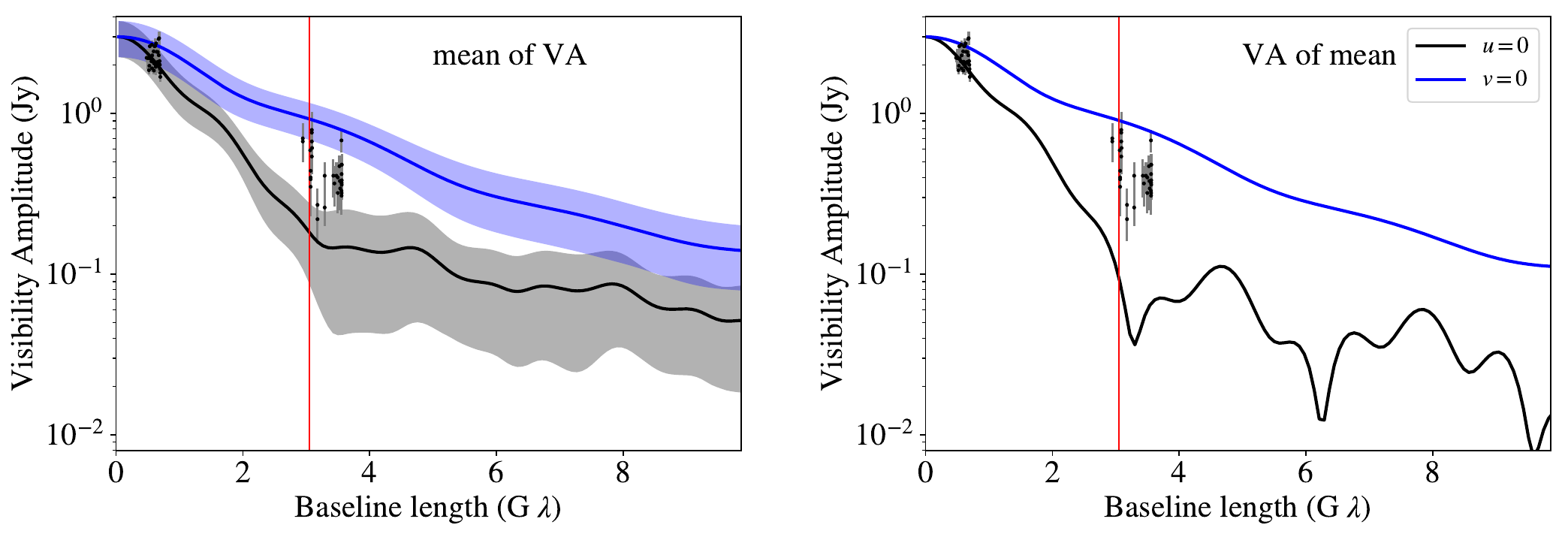}
\caption{(Top Left) Mean simulated image of Model A. Red circle indicates the location of the black hole shadow.
(Top Middle) Mean of the visibility amplitude of each snapshot. Red circle indicates the first null in the visibility amplitude of a thin photon ring located at the radius of the black hole shadow.
(Top Right) Visibility amplitude of the mean simulated image. Red circle is the same as the top middle panel.
(Bottom Left) Cross sections, taken parallel (black line and region) and perpendicular (blue line and region) to the black hole spin axis, of the top middle panel. These cross sections were not chosen to correspond to any particular EHT baselines. The colored regions are the $68\%$ ranges of the mean visibilities at each baseline. 
Red line is the location of first null in the visibility amplitude of a thin photon ring located at the radius of the black hole shadow.
(Bottom Right) Cross sections, taken parallel (black line) and perpendicular (blue line) to the black hole spin axis, of the top right panel. The red line is the same as the bottom left panel.
The black points and error bars in the bottom panels are EHT data taken in 2007 and 2009 shown here for illustrative purposes only (\citealt{2008Natur.455...78D}, \citealt{2011ApJ...727L..36F}).
\\
\\ }
\label{fig:snapA}
\end{figure*}

\begin{figure*}[t!]
\centering
\includegraphics[height=2in]{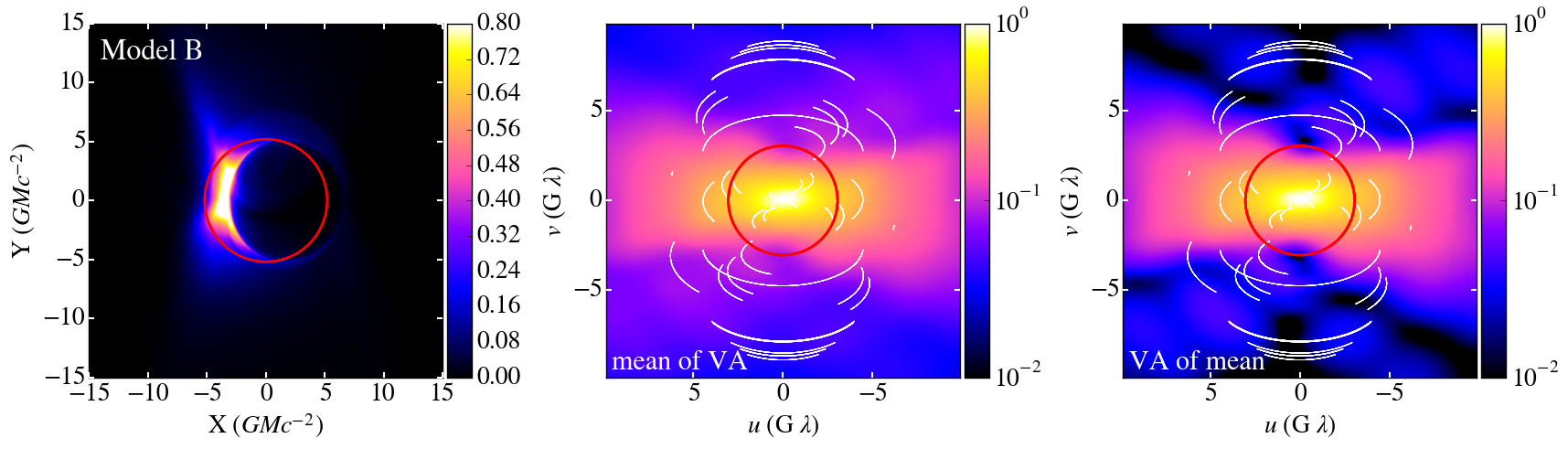}
\includegraphics[height=2in]{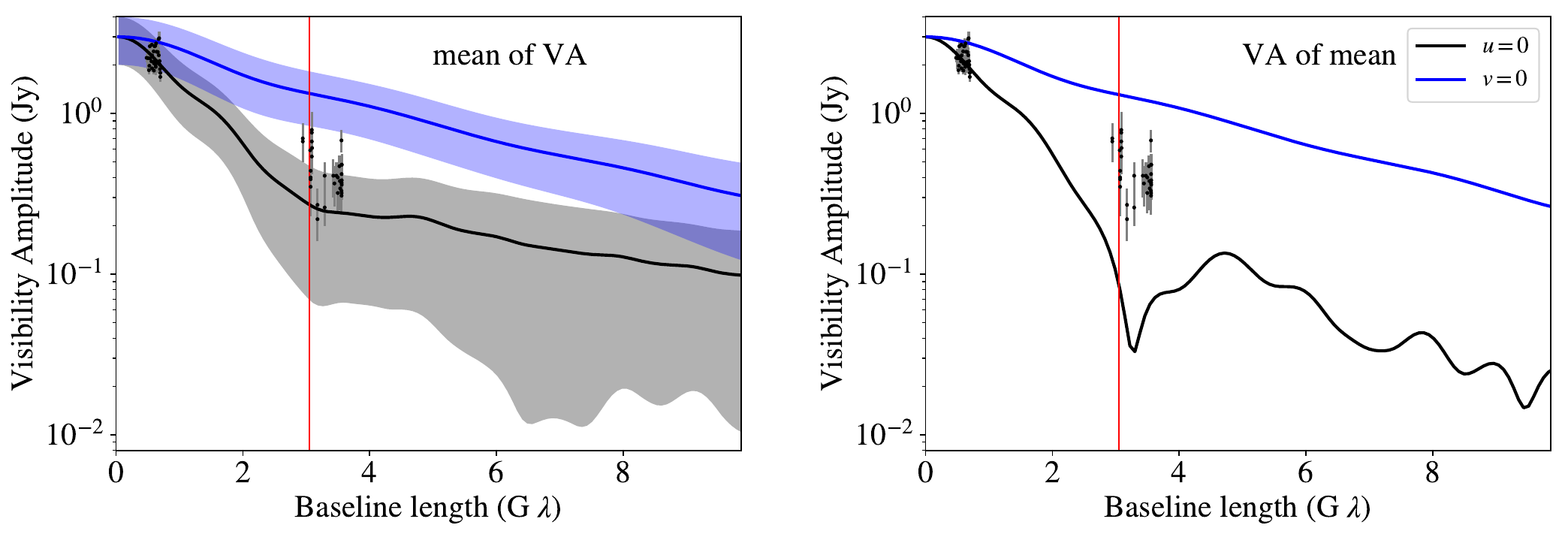}
\caption{Same as Figure~\ref{fig:snapA}, but for Model B.}
\label{fig:snapB}
\end{figure*}

\subsection{SANE Models}
The SANE models A and B have their 1.3~mm emission structure dominated by the disk and have  crescent-like shapes.
In Figures~\ref{fig:snapA} and \ref{fig:snapB}, we show the mean simulated images, the means of the visibility amplitudes of each snapshot, and the visibility amplitudes of the mean simulated images, for these SANE models.
Note that the visibility amplitude is the magnitude of the Fourier transform of the image. 
Because calculating the amplitude is a non-linear operation, the order of operations for taking the mean and calculating the amplitude matters.
Comparing the mean visibility amplitudes to the visibilities of the mean images reveals an expected but important consequence of variability.
The visibilities of any snapshot (including of the average image) have significantly more structure than the average visibilities.

\begin{figure*}[t!]
\centering
\includegraphics[width=\textwidth]{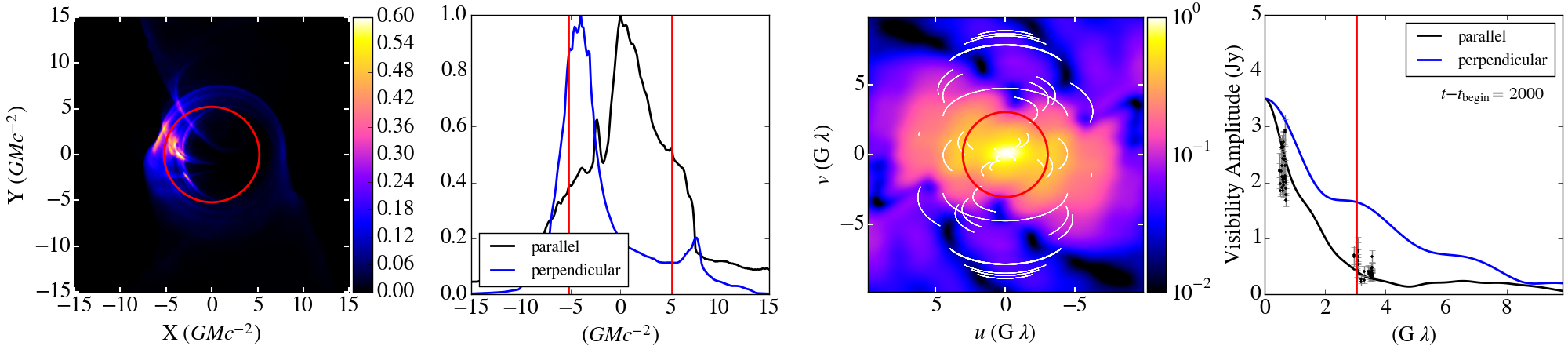}
\includegraphics[width=\textwidth]{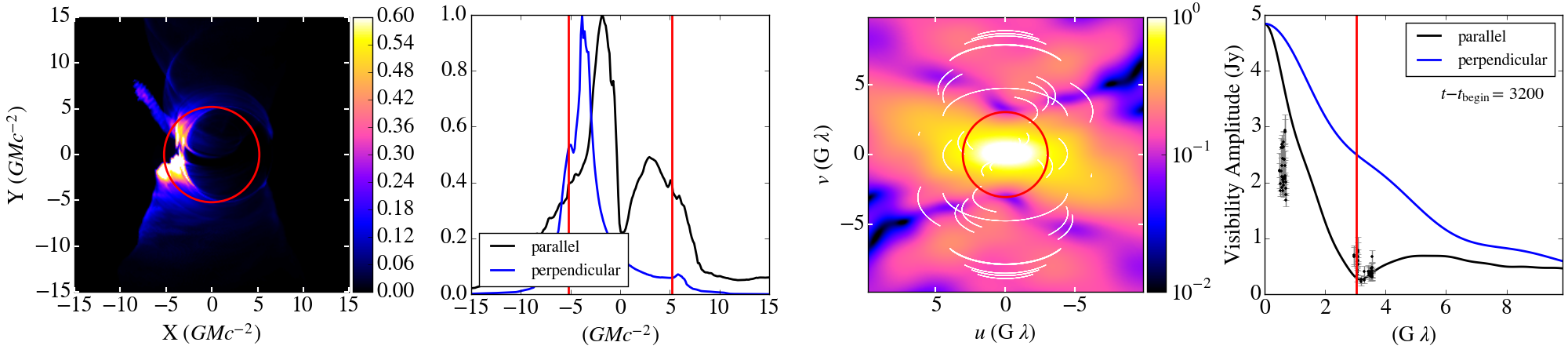}
\includegraphics[width=\textwidth]{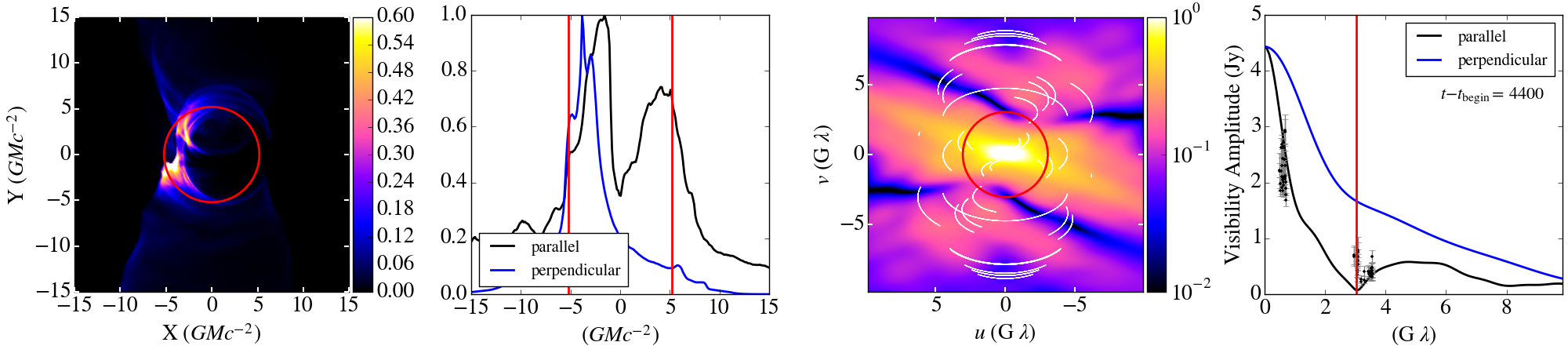}
\includegraphics[width=\textwidth]{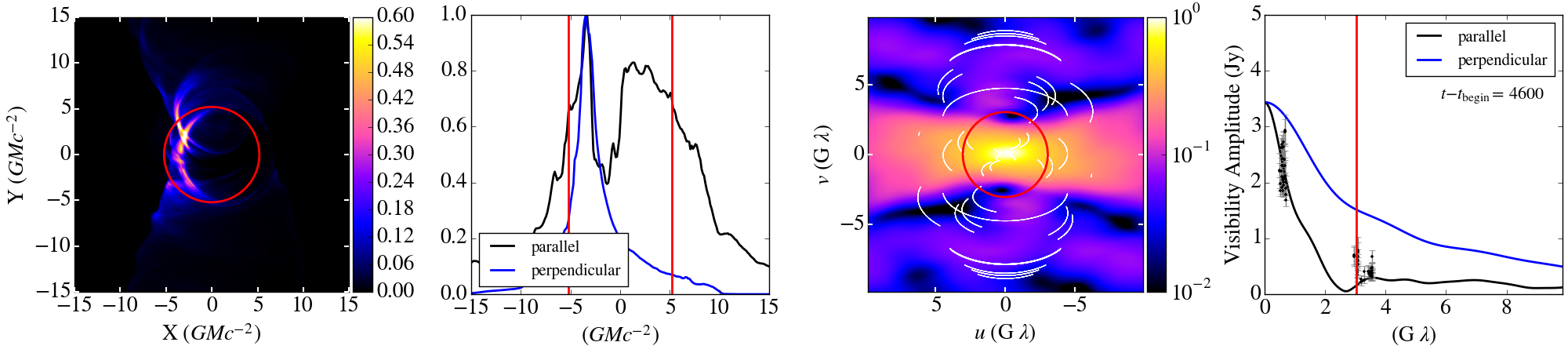}
\includegraphics[width=\textwidth]{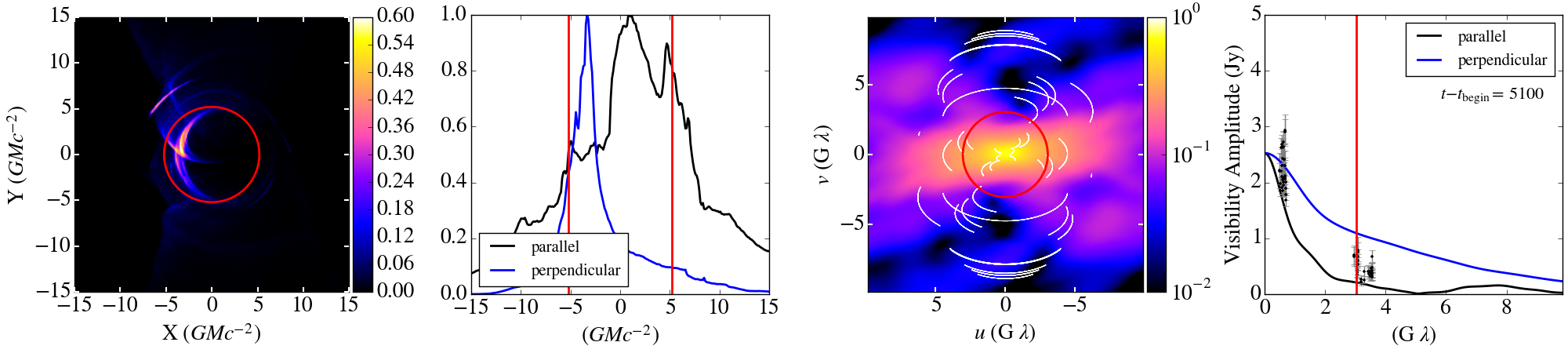}
\caption{Example snapshots from Model B. From left to right: the simulated image for each snapshot, the projections of the simulated image in directions parallel and perpendicular to the black hole spins axis, the visibility amplitude for this snapshot, and the cross sections for the visibility amplitude of this snapshot. }
\label{fig:snapsB}
\end{figure*}

To explore the behavior of the structure of visibility amplitudes further, we turn our focus to the images from the individual snapshots.
In the various panels of Figure~\ref{fig:snapsB}, we show (a) the simulated images, (b) the projections of these images along directions parallel and perpendicular to the black hole spin axis, (c) the visibility amplitudes, and (d) the cross sections of the visibility amplitudes taken along directions parallel and perpendicular to the black hole spin axis for 5 snapshots from model B.
The projection-slice theorem states that the Fourier transform of the projection of a two-dimensional image onto some axis is equal to a slice of the Fourier transform of the image as long as the slice is parallel to the projection axis and intersects the center of the visibility amplitude map. 
As a result, the cross sections of the visibility amplitudes shown in the rightmost panels are just the one-dimensional Fourier transforms of the projections shown in the second column of panels.
These cross sections are representative of the range of behavior of the two dimensional visibilities since the parallel cross section probes the closest deep minimum to the zero baseline while the visibility amplitudes are smoothly decreasing in the perpendicular cross section.
The cross sections are good representations for the range of behavior that could be observed for different orientations. 

In the rightmost column, we see that the visibility amplitude cross sections perpendicular to the spin axis decrease slowly and smoothly while those parallel to the spin axis have a lot more structure.
Particularly, the cross sections parallel to the spin axis often have minima, but the location and depth of these minima are variable.
By taking the average of the visibility amplitudes, we lose all information about the minima and are left with a smoothly decreasing visibility amplitude, as was seen in the mean of the visibilities in Figures \ref{fig:snapA} and \ref{fig:snapB}.

\begin{figure}[t!]
\centering
\includegraphics[height=2.5in]{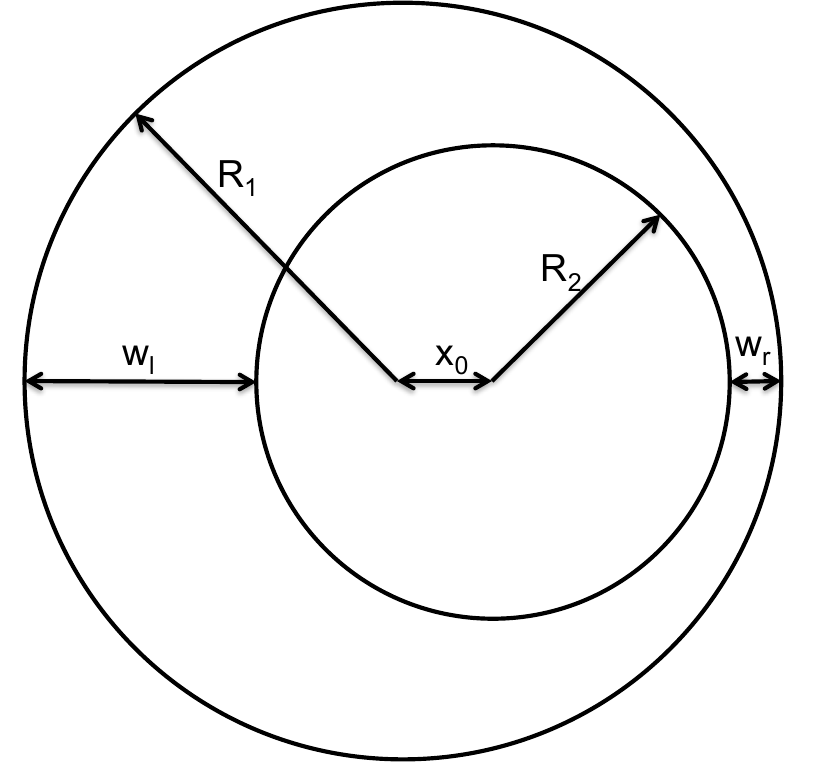}
\caption{An analytic representation of an asymmetric ring model (following \citealt{2013MNRAS.434..765K}) as a difference between two offset disks with different radii.}
\label{fig:cres}
\end{figure}

\begin{figure}[t!]
\centering
\includegraphics[height=3in]{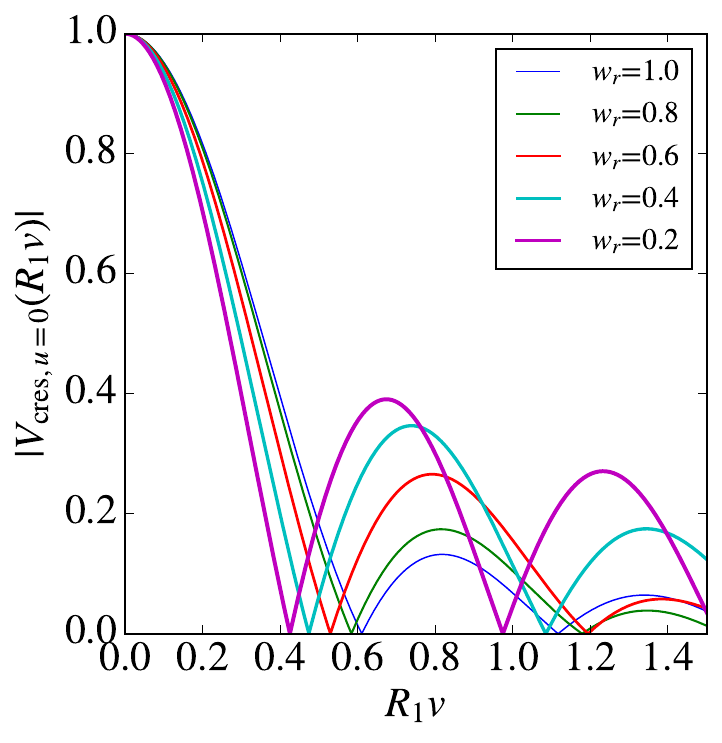}
\caption{The cross section of the visibility amplitude which is parallel to the spin axis for the analytic asymmetric ring model for different widths of the ring. 
Since the parallel cross section does not depend on the asymmetry of the ring, we have set $w_l=w_r$.
A characteristic null appears in the visibility amplitude at $v\approx0.5/R_1$, with its precise location depending on the width of the ring. }
\label{fig:cresR}
\end{figure}

To understand the behavior described above, we employ a simple analytic model to represent the properties of the emission regions.
Since the SANE models (A and B) have their emission dominated by the Doppler boosted disk, it roughly resembles a crescent shape.  
Following \citet{2013MNRAS.434..765K}, we use a model of an asymmetric ring, defined as the difference between two offset disks with different radii, which becomes a crescent as its asymmetry grows.
The diagram in Figure~\ref{fig:cres} shows the parameters used to describe our asymmetric ring model.
The Fourier transform of the asymmetric ring is given by
\begin{equation}
V_{\mathrm{cres}}(k) =  \frac{2 \pi I_0 R_1 }{k}\left[J_1\left(R_1 k\right)  -e^{-2\pi i(\alpha_0 u)} RJ_1\left(kR_2\right)    \right],
\end{equation}
where $k\equiv 2\pi \sqrt{u^2+v^2}$, $J_1$ are Bessel functions of the first kind, $I_0$ is the constant surface brightness of the disks, $\alpha_0$ is the displacement of the smaller disk from the center of the larger disk in the $\alpha$ direction, and $R=R_2/R_1$. 
Hereafter, we set $R_1=1$ without loss of generality.
We also define the widths of the asymmetric ring on the left and right sides of the image as $w_l\equiv R_1-R_2+\alpha_0$ and $w_r\equiv R_1-R_2-\alpha_0$.

A cross section of the visibilities parallel to the black hole spin axis is then given by
\begin{equation}
V_{\mathrm{cres},u=0}(v) =  \frac{I_0 }{v}\left[J_1\left(2\pi v\right)  -RJ_1\left(2\pi Rv \right) \right],
\end{equation}
which only depends on $R$, the ratio of the two radii, and not on $\alpha_0$, the displacement of the smaller disk. 
In other words, the visibility amplitude along the directions parallel to the spin axis will be the same regardless of the asymmetry of the ring.
For an infinitesimally thin ring, i.e., when $w_l\ll R_1$ and $w_r\ll R_1$, the visibility amplitude along a direction parallel to the spin axis has a minimum at $u\simeq 0.4/R_1$. 
However, Figure~\ref{fig:cresR} shows that changing the width of the ring, $w_l=w_r=R_1-R_2$, changes the location of the minima.
In the SANE simulations (A and B) the width of the approximately crescent shape is highly variable because of different turbulent structures appearing and disappearing from the Doppler boosted side of the crescent.
This is why the locations of the minima are also highly variable. 

\begin{figure}[t!]
\centering
\includegraphics[height=3in]{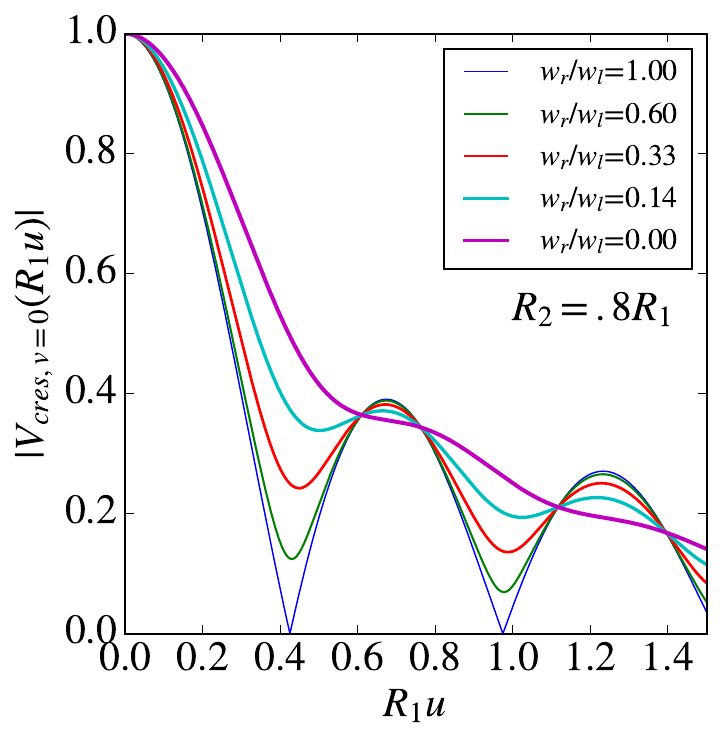}
\caption{The cross section of the visibility amplitude in the direction perpendicular to the spin axis for the asymmetric ring model, for different degrees of asymmetry in the ring brightness.
As the ring becomes more asymmetric, the local minimum at $u\approx0.5/R_1$ becomes less pronounced.}
\label{fig:cresx}
\end{figure}

The cross section of the visibility function perpendicular to the spin axis of the black hole ($v=0$) is equal to
\begin{equation}
V_{\mathrm{cres},v=0}(u) =  \frac{I_0 }{u}\left[J_1\left(2\pi u\right)  -e^{-2\pi i \alpha_0 u} RJ_1\left(2\pi Ru \right) \right].
\end{equation}
Although the visibility amplitude of a cross section parallel to the spin axis did not depend on the displacement $\alpha_0$, which measures the asymmetry of the ring, the perpendicular cross section does. 
Figure~\ref{fig:cresx} shows the dependence of the visibility amplitude on the degree of asymmetry of the ring $w_r/w_l$.
As the ring becomes more asymmetric, i.e., as $w_r/w_l \rightarrow 0$, the visibility amplitude of the cross section perpendicular to the spin axis becomes smoother and all traces of minima are lost.
Due to the effects of Doppler beaming, our simulations are completely dominated by one side of the ring. 
This is why the cross sections of the visibilities perpendicular to the spin axis (e.g., the blue curves in Figures \ref{fig:snapA} and \ref{fig:snapB}) are broad and monotonically decreasing, while the parallel cross sections have strong local minima.

The model described above cannot fully encompass all of the variable structure that we see in Figure~\ref{fig:snapsB}.
In the GRMHD simulations, the emission is not always a simple crescent. In some snapshots, the emission along the equatorial plane gets blocked by the colder disk which causes the emission to have two disjointed regions.
When this occurs, the visibility amplitudes have short-lived features similar to those of the MAD models, which have two disjointed peaks of emission and will be discussed below.

In summary, for most instances, the behavior of the SANE models (A and B) can be roughly modeled by an asymmetric ring of variable width. 
The behavior of the cross section of the visibility amplitude that is parallel to the spin axis does not depend on the asymmetry of the ring and exhibits minima. 
The location of these minima depends on the width of the asymmetric ring, which is variable. 
Because of this variability, taking the average of the visibility amplitude over time will result in a visibility amplitude with reduced or no minima. 
The direction perpendicular to the spin axis, however, does depend on $\alpha_0$.
An extremely offset ring, where the thinnest part has a thickness of zero, has a monotonically decreasing visibility amplitude. 

\begin{figure*}[t!]
\centering
\includegraphics[height=2in]{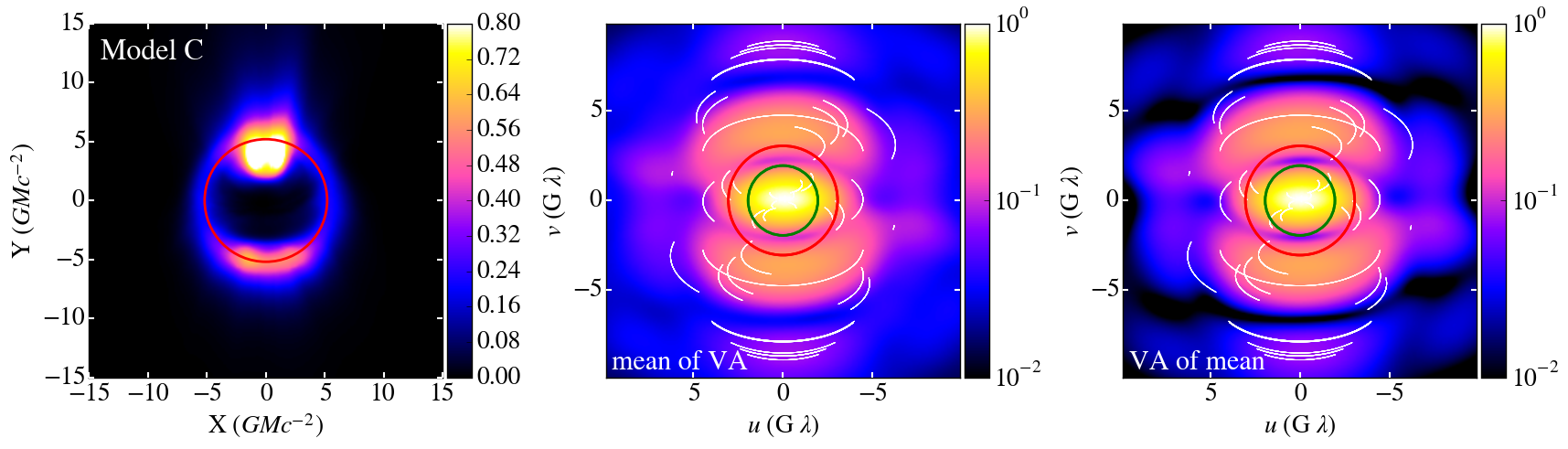}
\includegraphics[height=2in]{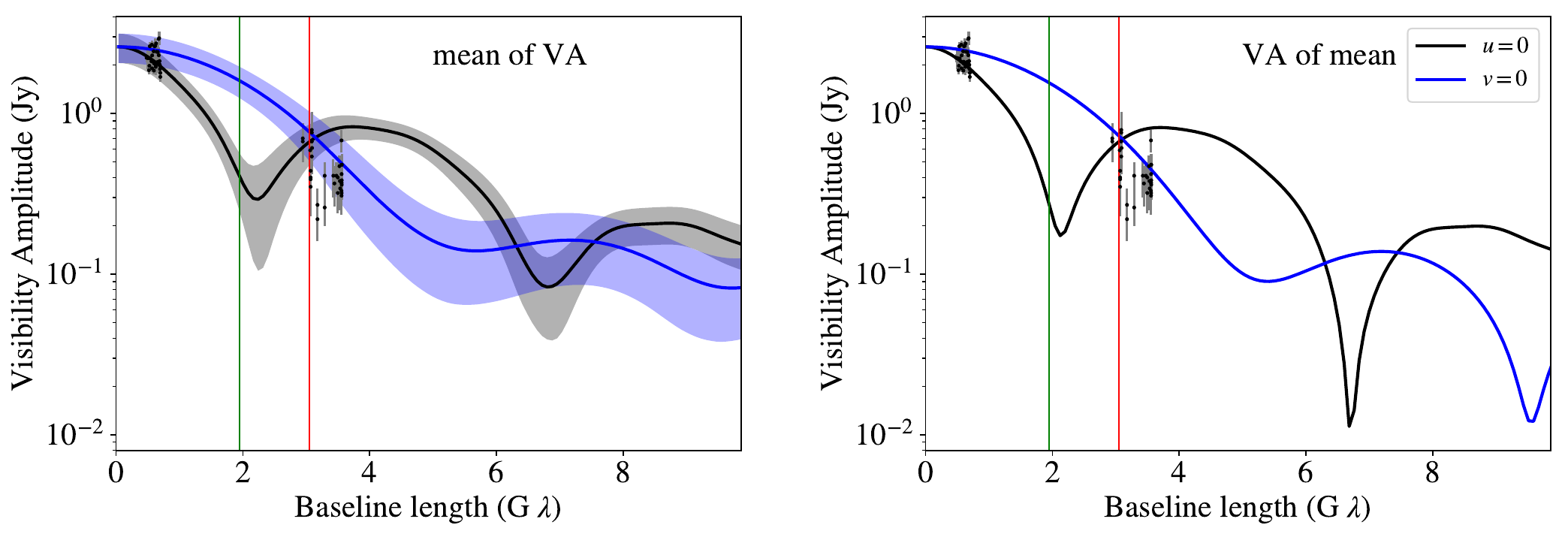}
\caption{Same as Figure~\ref{fig:snapA}, but for the images from Model C and with the addition of green circles in the top middle and right panels and green lines in the bottom panels. The location of these corresponds to the location of the first minimum in the visibility amplitude of two Gaussians separated by a distance equal to the size of the black hole shadow. These minima occur at small baseline lengths compared to the minima of the asymmetric ring shown in red.
}
\label{fig:snapC}
\end{figure*}
\subsection{MAD Models}

\begin{figure*}[t!]
\centering
\includegraphics[height=2in]{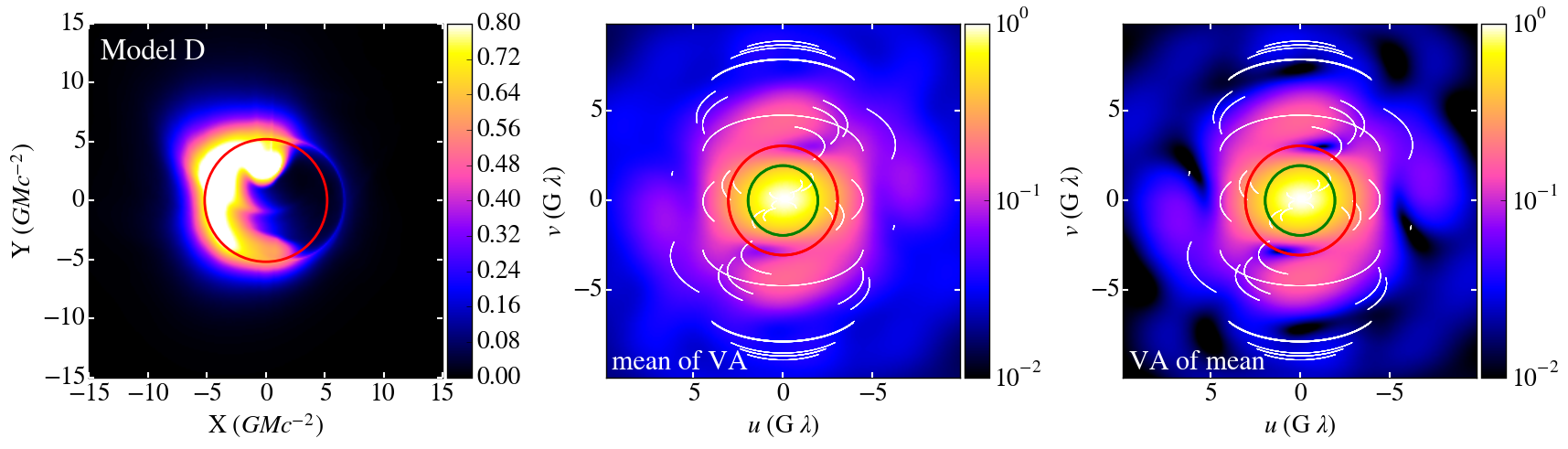}
\includegraphics[height=2in]{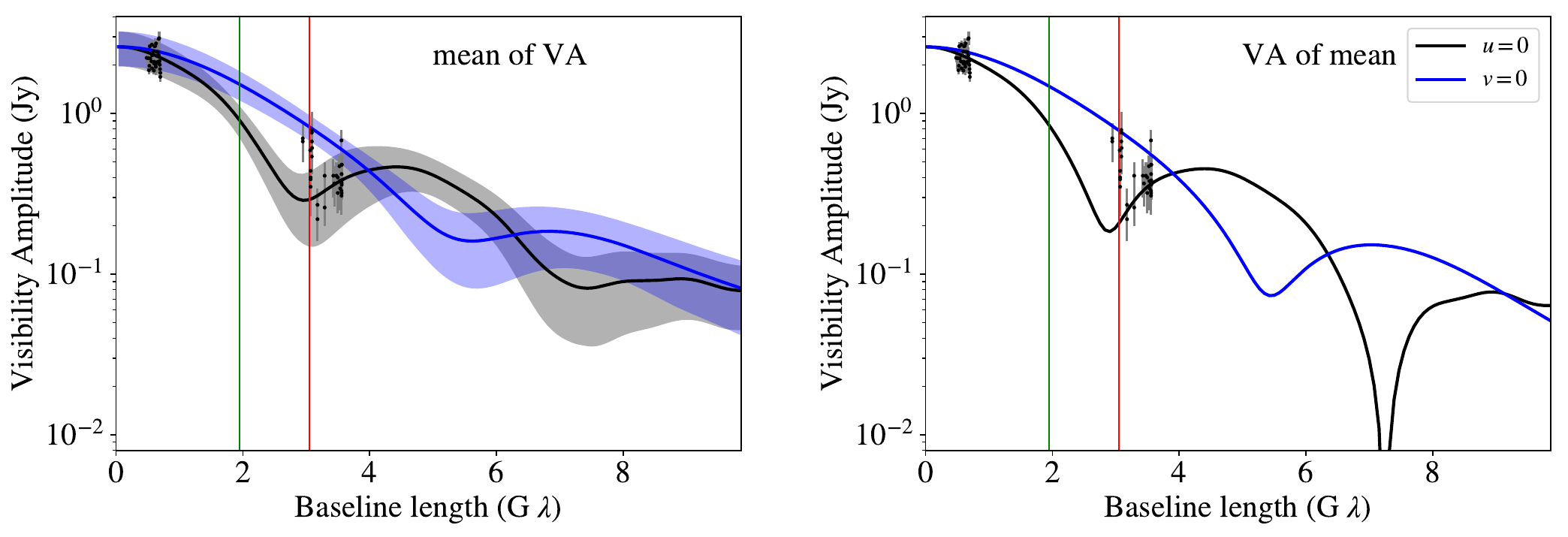}
\caption{Same as Figure~\ref{fig:snapC}, but for Model D.
}
\label{fig:snapD}
\end{figure*}

The 1.3~mm emission of the MAD models (C, D, and E) is dominated by the funnels and jet footprints; because of this, it is characterized broadly by two peaks. 
Throughout the simulations, the relative widths and amplitudes of the peaks change but the distance between them remains approximately constant since it is set by the size of the black hole shadow. 

Figures \ref{fig:snapC} and \ref{fig:snapD} are the equivalent of Figure~\ref{fig:snapA} but for Models C and  D, respectively.
Focusing on the one-dimensional cross sections of the visibility amplitude of the mean and the mean of the visibility amplitudes, the cross sections perpendicular to the spin axis have less pronounced minima, which are located at baselines much larger than expected for the size of Sgr~A$^*$.
The cross sections parallel to the spin axis show much more pronounced minima, close to the expected location for the size of Sgr~A$^*$, much like the SANE models.
Unlike the SANE models, however, the means of the visibility amplitudes and the visibility amplitudes of the means have very similar structures and both have a minimum in the vertical direction.
Furthermore, the minima appear to be in the same locations and the ranges of amplitudes at a given baseline are much smaller than those of the SANE models.
This indicates that the existence and location of a visibility minimum is more persistent in MAD models than in SANE models. 

\begin{figure}[t!]
\centering
\includegraphics[height=2.5in]{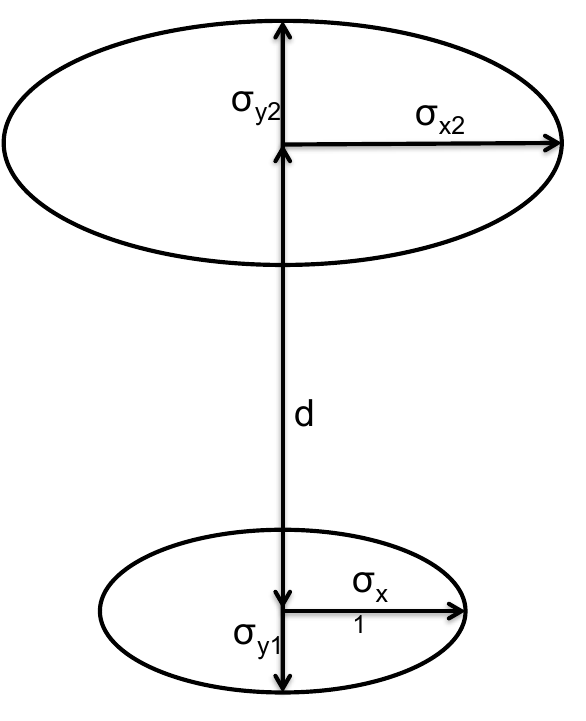}
\caption{An analytic representation of a two spot model in terms of two Gaussian components separated by a distance $d$ along the vertical direction. }
\label{fig:2gauss}
\end{figure}

To understand the behavior of the MAD models, we employ an approximate model of their emission using two Gaussians separated by a distance $d$ in the direction parallel to the spin axis but with no separation in the direction perpendicular to the spin axis. 
We define this model as (see Figure~\ref{fig:2gauss})
\begin{eqnarray}
I(\alpha)=&&A_1 e^{-\left[(\alpha-\alpha_{01})^2/2 \sigma_{\alpha1}^2+ (\beta-\beta_{01})^2/2 \sigma_{\beta1}^2\right]}\nonumber \\
&&+ A_2 e^{-\left[(\alpha-\alpha_{02})^2/2 \sigma_{\alpha2}^2 + (\beta-\beta_{02})^2/2 \sigma_{\beta2}^2\right]}.
\end{eqnarray}
To simplify our notation, we set $\alpha_{01}=\beta_{01}=0$ such that one Gaussian is peaked at the origin; set $\alpha_{02}=0$ so that the Gaussians are separated by a distance $d=\beta_{02}$ along the $\beta$ axis, and define the ratio of amplitudes $A \equiv A_2/A_1$. This reduces to an overall normalization of $A_1$, which we ignore to get
\begin{equation}
I(\alpha)= e^{-\left[\alpha^2/2 \sigma_{\alpha1}^2+ \beta^2/2 \sigma_{\beta1}^2\right]} + A e^{-\left[\alpha^2/2 \sigma_{\alpha2}^2 + (\beta-d)^2/2 \sigma_{\beta2}^2\right]}.
\end{equation}
We further define the quantities 
\begin{eqnarray}
\alpha'&&\equiv\frac{\alpha}{\sigma_{\alpha1}}, \,\,\,\,\,\,    \sigma_{\alpha}'\equiv\frac{\sigma_{\alpha2}}{\sigma_{\alpha1}}, \,\,\,\,\,\,     \beta'\equiv\frac{\beta}{\sigma_{\beta1}}, \nonumber\\
\sigma_{\beta}'
&&\equiv\frac{\sigma_{\beta2}}{\sigma_{\beta1}}, \,\,\,\,\,\,\ d'\equiv\frac{d}{\sigma_{\beta1}}, 
\end{eqnarray}
which in turn gives 
\begin{eqnarray}
I(\alpha)&&=e^{-\left(\alpha'^2 +\beta'^2\right)/2} + Ae^{-\left(\alpha'^2/\sigma_{\alpha}' + (\beta'-d')^2/\sigma_{\beta}'^2\right)/2}\nonumber\\
&&=e^{-\left(\alpha^2 +\beta^2\right)/2} + Ae^{-\left(\alpha^2/\sigma_{\alpha} + (\beta-d)^2/\sigma_{\beta}^2\right)/2}.
\end{eqnarray}
In this last expression, we omitted the primes for clarity.
We now take the Fourier transform of the intensity to obtain
\begin{eqnarray}
V_{2\mathrm{g}}(k) = &&2\pi e^{-\left(u^2+v^2\right)2\pi^2}\nonumber\\
&&+2\pi A\sigma_{\alpha}\sigma_{\beta} e^{-\left(u^2\sigma_{\alpha}^2 +v^2\sigma_{\beta}^2\right)2\pi^2}e^{-2\pi idv}.
\end{eqnarray}

The visibility along a direction parallel to the spin axis takes the form
\begin{equation}
V_{2\mathrm{g}, u=0}(v) = 2\pi e^{-2(\pi v)^2}+2\pi A\sigma_{\alpha}\sigma_{\beta} e^{-2(\pi \sigma_{\beta} v)^2}e^{-2\pi idv}.
\end{equation}
Its magnitude, $| V_{2\mathrm{g}}(v) |$, has a minimum when $V_{2\mathrm{g}}(v)$ is minimum, i.e., when the second term is real and negative.
This occurs when 
\begin{equation}
 vd= (2n+1)/2, 
 \label{eq:nulld}
 \end{equation}
 i.e., the location of the minimum depends only on the separation between the two Gaussians and not on any of their other properties. 
 Note that, in this configuration, the location of the minimum occurs at a baseline length that is different compared to the case of an asymmetric ring model (see Figures \ref{fig:snapC} and \ref{fig:snapD}).
This minimum will reach zero when the amplitudes of the two terms in the sum are equal, or when
\begin{equation}
A = \frac{1}{\sigma_{\alpha}\sigma_{\beta}} e^{-\pi^2 (1-\sigma_{\beta}^2)(2n+1)^2/2d^2}.
\end{equation}

\begin{figure}[t!]
\centering
\includegraphics[height=3in]{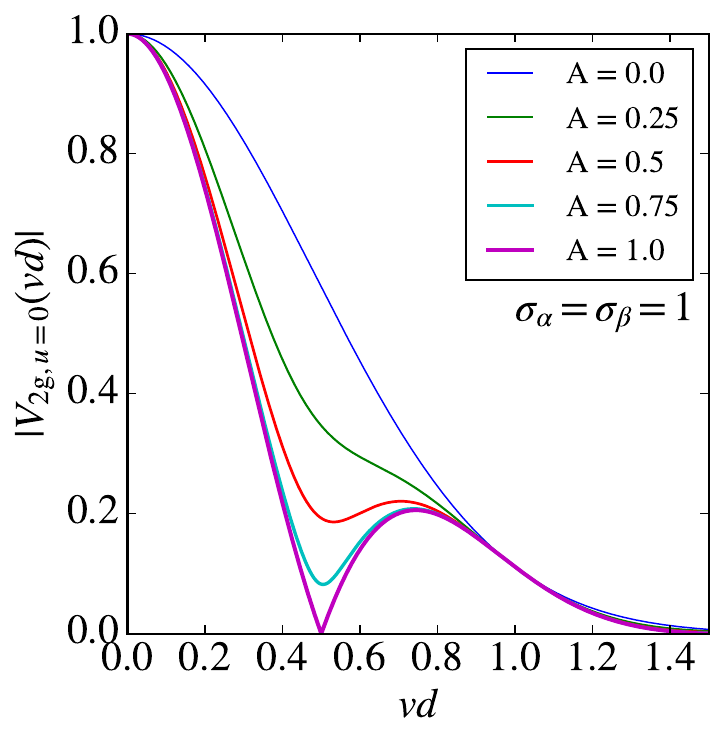}
\caption{The cross section of the visibility amplitude in the direction parallel to the spin axis, for the analytic two-component model for different relative brightnesses of the components, $A$.
We have also set the displacement to $d=3$ and the relative widths to $\sigma_{\alpha}=\sigma_{\beta}=1$.
The baseline dependence of the visibility amplitude shows a characteristic minimum at a location that depends only on the distance between the two components of the image. 
The minimum becomes deeper as the relative brightness of the two components becomes equal to unity.}
\label{fig:2gaussA}
\end{figure}

\begin{figure}[t!]
\centering
\includegraphics[height=3in]{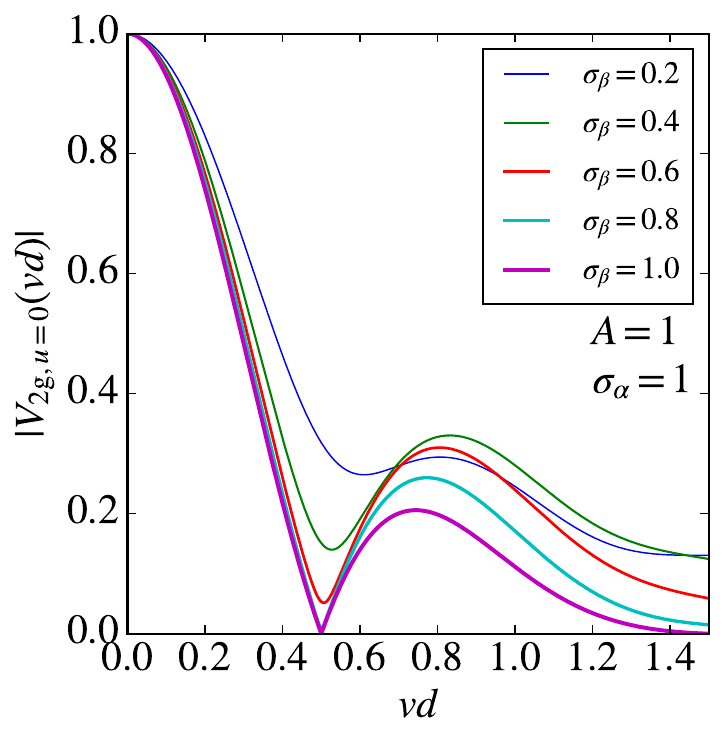}
\caption{The cross section of the visibility amplitude in the direction parallel to the spin axis, for the analytic two-component model for different widths of the components, $\sigma_{\beta}$. 
We have also set the displacement to $d=3$, the relative widths in the $\alpha$ direction to $\sigma_{\alpha}=1$, and the relative brightnesses to $A=1$.
As in Figure~\ref{fig:2gaussA}, the visibility amplitudes show a characteristic minimum around $v=0.5/d$. 
The minimum becomes deeper as the relative width of the two components of the image becomes one. }
\label{fig:2gaussS}
\end{figure}

The simplified analytic model shows that two Gaussians do not always produce a null in the visibility amplitude. 
For a given separation $d$ and width ratios $\sigma_{\alpha}$ and $\sigma_{\beta}$, there is only one value of the ratio of the brightness of the two Gaussians, $A$, that gives rise to a null. 
When a local minimum does not reach zero, the properties of the two components of the image affect the depth of the minimum.
This is shown in Figure~\ref{fig:2gaussA} for the dependence of the depth of the minimum on the ratio $A$ of the brightness of the two components of the image and, in Figure~\ref{fig:2gaussS}, for the dependence of the ratio of widths of the two components along the spin axis of the black hole.
In our MAD simulations, the relative amplitudes and widths of the two peaks is highly variable; however, the distance between the two peaks remains approximately constant and is set by the size of the black hole shadow and the observer inclination.
Because of this, the depth of the minimum varies but the location is approximately constant. 
When we average minima of various depths but constant location, we get a minimum in the same location of an average depth.

The cross sections of the visibility amplitude maps perpendicular to the spin axis do not exhibit nulls. 
Analytically, the perpendicular cross sections of the visibility amplitudes of two Gaussians separated in the vertical direction is given by
\begin{equation}
V_{2\mathrm{g}, v=0}(u) = 2\pi e^{-2(\pi u)^2}+2\pi A\sigma_{\alpha}\sigma_{\beta} e^{-2(\pi \sigma_{\alpha} u)^2},
\end{equation}
 which is just the addition of two Gaussians both peaked at the origin, and therefore has no nulls.
 
In our simulations, the visibility amplitudes of the MAD models have a consistent minimum in the direction parallel to the spin axis of the black hole. 
The location of the minimum is approximately constant and is determined primarily by the size of the black hole shadow. Averaging over time does not appear to erase the minimum as it did in the SANE models. 

\section{Conclusions}
In this paper, we quantified the effect of MHD turbulence driven variability on the structure of visibility amplitudes of the upcoming imaging observations of Sgr~A$^*$ with the EHT.
We explored the effect of variability on the structure of the emission region in order to understand the challenges that variability will pose for image reconstruction of EHT observations. 
We created and analyzed mock images and $u-v$ maps using GRMHD simulations, that were constrained in previous work such that their time averaged broadband spectrum and 1.3~mm image size match observations of Sgr~A$^*$. 

We found that the visibility amplitude of the SANE models resembles that of a highly asymmetric ring.
The width of the asymmetric ring is highly variable due to the turbulent accretion flow. 
The visibility amplitude in the direction parallel to the spin axis of the black hole typically exhibits minima with locations that depend on the width of the asymmetric ring. 
Since the location of the minima in the direction parallel to the spin axis depends on the width of the emitting region, and is therefore variable, any information that could be inferred by the presence of a minimum is lost by averaging over time.

The SANE models rarely exhibit minima in the direction perpendicular to the spin axis of the black hole.
The reason for this is that, due to Doppler beaming, the majority of emission comes from the left of the spin axis (for the spin orientation we use in our figures), with negligible emission coming from the right.
This asymmetry does not affect the visibility amplitude in the direction parallel to the spin axis, but affects the depth of minima in the direction perpendicular to the spin axis.
For the perpendicular direction, a highly asymmetric ring has a visibility amplitude that decreases monotonically. 

In contrast, the images and visibility amplitudes of the MAD models are characterized by two bright spots at the footpoints of the jets, separated by a relatively constant distance equal to the size of the black hole shadow. 
For the MAD models, the visibility amplitudes in the direction parallel to the spin axis have persistent nulls in constant locations but with variable depths.
The locations of the minima in the direction parallel to the spin axis of MAD models depend strongly on the separation between the two image components.
Since the distance between the emission peaks in our simulations are set primarily by the size of the black hole shadow and are approximately constant, the location of the minimum is constant.
However, varying the widths, or amplitudes, of the image components affects the depths of the minima.
On the other hand, the visibility amplitudes in the direction perpendicular to the spin axis have much less pronounced minima at larger baselines.

\begin{figure*}[t!]
\centering
\includegraphics[height=8.1in]{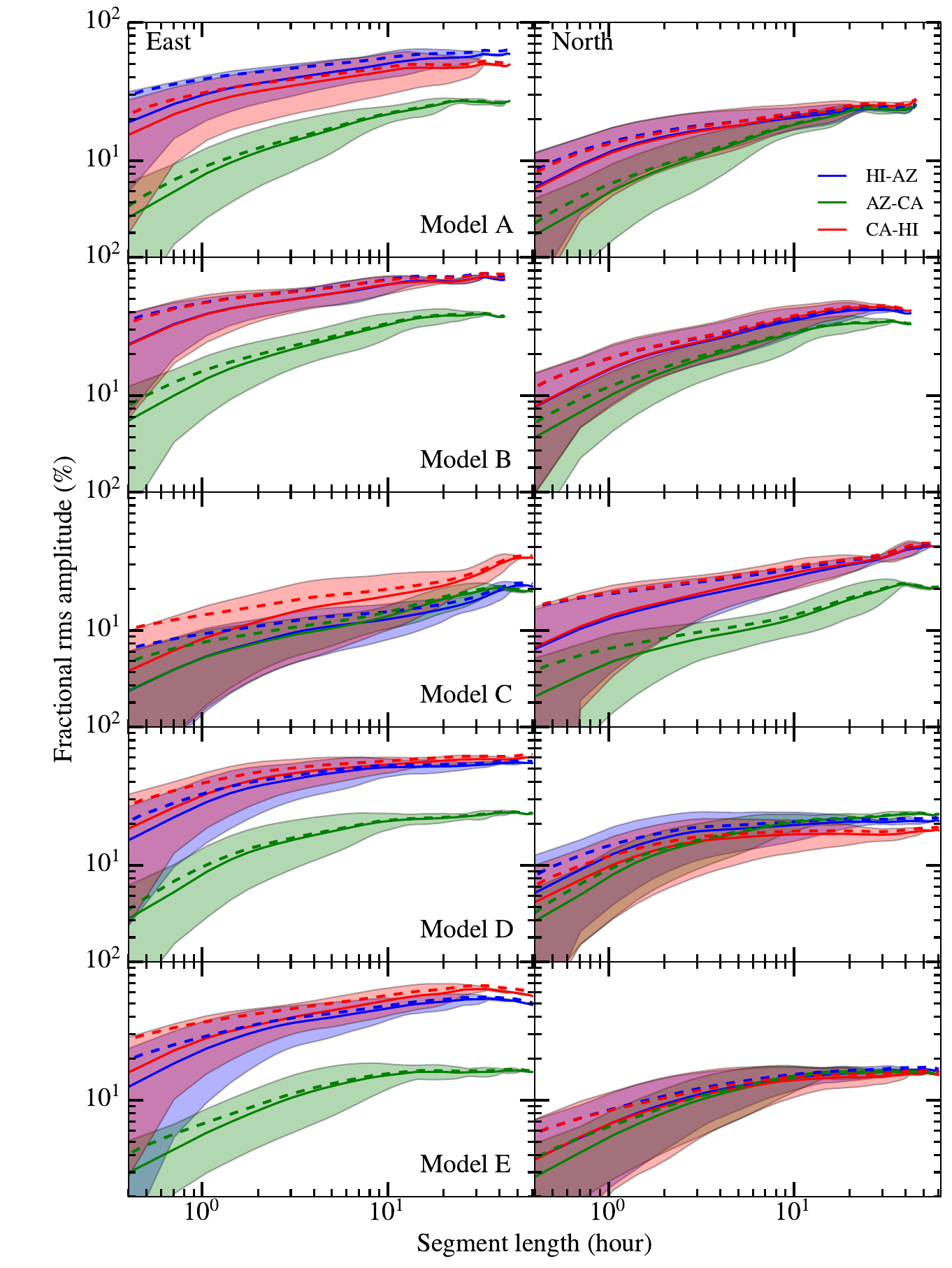}
\caption{Fractional root-mean-squared (rms) variability as a function of the time length of simulation segments. The right (left) column assumes that the  black hole spin axis points North (East). The dashed line in each panel is the mean fractional standard deviation of the original simulation with a time resolution of about 3.5 minutes and an effectively instantaneous exposure. The solid line is the mean fractional
standard deviation calculated using an effective exposure (averaging) time of 10.5 minutes.
The colored regions correspond to the standard deviation in the fractional rms variability for different overlapping segments in the simulation. The three colors in the plot correspond to the three baselines used by \citet{2008Natur.455...78D} and \citet{2011ApJ...727L..36F} to take the data shown in Figures \ref{fig:snapA}, \ref{fig:snapB}, \ref{fig:snapsB}, \ref{fig:snapC}, and \ref{fig:snapD}. Red corresponds to the baseline formed by CARMA (California) and the SubMillimeter Telescope (SMT, Arizona), green corresponds to the baseline formed by CARMA (California) and the sub millimeter array (SMA, Hawaii), and blue corresponds to SMT (Arizona) and SMA (Hawaii). Segments that are at least a few hours long exhibit the full range of variability that our simulations show.
\\
\\}
\label{fig:frac_rms}
\end{figure*}

As discussed in the introduction, typical EHT exposure times are around 10 minutes and typical imaging runs span a few hours. Our 60-hour simulations are longer than a typical observing run but shorter than the timespan between different observing epochs. Indeed, the EHT is expected to observe Sgr~A$^*$ for a few nights in each observing cycle, with multiple cycles to occur in successive years. Therefore, the EHT data set as a whole will span multiple years and will sample a broad range of the variability of the source. If the longest timescale of significant source variability is of the order of a few hours to a day, then the range of variability that our models exhibit will be representative of the expected range of variability in the data, when multiple epochs are combined. 

A second important issue that we can address with our simulations is the effect of the length of each imaging run on the degree of expected variability. To achieve this, we divided each one of our simulations into overlapping segments of constant length and calculated the fractional rms variability for each segment of the visibility amplitude at three locations in the $u-v$ plane that correspond to the Arizona-California-Hawaii baselines. We then use the measured values of rms variability and calculate their average and standard deviation. In all cases, we calculate the fractional variability by dividing the standard deviation with the mean visibility amplitude of the entire 60~hour simulation. Finally, we explored the effect on the fractional rms variability of using different exposure (i.e., averaging) times for each measurement of a visibility amplitude.

In Figure \ref{fig:frac_rms}, we show the dependence of these fractional rms variability amplitudes on the length of the segments. For displaying the mean of the rms variability, we use dashed and solid lines to show the effect of changing the exposure time from the 3.5~minute cadence of our simulation outputs to a 10.5~minute time that is representative of early EHT observations. Changing the effective exposure time has a very little impact on the rms variability. On the other hand, as the length of each segment increases, both the number of different overlapping segments decreases and the amount of overlap between successive segments increases, as well. Because of this, the range of fractional rms variability decreases as the length of each segment increases. 

In all five models, the fractional rms amplitude increases rapidly with the length of the segment over which the variability is calculated, for lengths shorter than a few hours. Beyond that, increasing the length of the segment does not contribute significantly to the variability. This is expected because, as we discussed in \citet{2015ApJ...799....1C}, the power spectrum of the 1.3~variability exhibited by our simulations is that of red noise and turns over at timescales longer than a few hours. This is also consistent with the reported turnover timescale for the observer power spectrum of Sgr A* at 1.3 mm \citep{2014MNRAS.442.2797D}. Our results suggest that observing runs that are a few hours long will be affected by the full range of variability that our simulations exhibit.

The variability in the interferometric visibilities may, in principle, affect both the parameter estimations based on comparing theoretical models to data and the images that will be reconstructed from the observations. In a companion paper \citep{2016arXiv160200692K}, we develop a Bayesian method to 
perform parameter estimation using explicitly the time dependence of GRMHD simulations and the expected variability of the upcoming EHT observables. Even without employing such a method, our results nevertheless suggest that fitting early EHT data using simple models of the accretion flow that do not take into account its intrinsic variability will lead to a reasonably accurate determination of the orientation of the black-hole spin on the plane of the sky that will depend only weakly on the flow properties (see, e.g., \citealt{2011ApJ...735..110B,2015ApJ...798...15P}).  This is true because, in both the disk-dominated SANE models and the
jet-dominated MAD models, the visibility amplitudes perpendicular to the spin axes are
smoother and vary over longer baselines while the visibilities along the spin axes have
significantly more structure and drop faster with baseline length.  

Using, on the other hand, simple time-independent models to measure more detailed properties of the black hole, such as the size of its shadow and the magnitude of its spin will be likely hampered by the variability in the flow because such models do not generally allow the image structure to change. For example, the spin of the black hole affects primarily the widths of the crescent-like shapes of disk-dominated images and, hence, the locations and depths of minima in the visibility amplitudes. However, turbulence in the accretion flow causes both the locations and
depths of these minima to be highly variable, effectively masking the effect of black-hole spin
on the image. In order for the EHT observations to lead to accurate determination of the black-hole shadow
size and spin, the image reconstruction and model fitting algorithms will need to take into account the variability of the underlying images explicitly.

\acknowledgements
We thank Ramesh Narayan for collaborative work that contributed to this paper and for helpful comments and suggestions on the manuscript. 
L.M. acknowledges support from NFS GRFP grant DGE~1144085.
C.K.C., F.O., and D.P. were partially supported by NASA/NSF TCAN award
NNX14AB48G and NSF grants AST~1108753 and AST~1312034.
J.K., and D.M. acknowledge support from NSF grant AST-1207752 and AST-1440254.
All ray tracing calculations were performed with the \texttt{El~Gato}
GPU cluster at the University of Arizona that is funded by NSF award
1228509.

\bibliography{ms}
\end{document}